\documentclass[pra,twocolumn,showkeys,superscriptaddress,groupedaddress]{revtex4}

\usepackage{amsmath,amsthm,amsfonts,amssymb,times,graphicx,color}
\usepackage{multirow}
\usepackage{cases} 
\usepackage{subfigure}
\usepackage{mathtools}  
\usepackage{url}

\usepackage{hyperref} 

\usepackage{amsmath,amsfonts,amssymb,bbm}

\newcommand{\couic}[1]{}

\newcommand{\C}{\mathbb{C}}

\newcommand{\R}{\mathbb{R}}

\newcommand{\Z}{\mathbb{Z}}

\newcommand{\Id}{\operatorname{Id}}
\newcommand{\trace}{\textrm{Tr}}

\newcommand{\ii}{\mathrm i}

\newcommand{\transp}{\mathsf{T}}
\newcommand{\diag}{\textrm{diag}}

\newtheorem{lemma}{Lemma}

\newtheorem{theo}{Theorem}

\begin{document}

\title[]{Quantum walking in curved spacetime}

\author{Pablo Arrighi}
\email{pablo.arrighi@univ-amu.fr}
\affiliation{Aix-Marseille Univ., LIF, F-13288 Marseille, France}

\author{Stefano Facchini}
\email{stefano.facchini@imag.fr}
\affiliation{Univ. Grenoble Alpes, LIG, F-38000 Grenoble, France}

\author{Marcelo Forets}
\email{marcelo.forets@imag.fr}
\affiliation{Univ. Grenoble Alpes, LIG, F-38000 Grenoble, France}

\date{\today}

\begin{abstract}
A discrete-time Quantum Walk (QW) is essentially a unitary operator driving the evolution of a single particle on the lattice. Some QWs admit a continuum limit, leading to familiar PDEs (e.g. the Dirac equation). In this paper, we study the continuum limit of a wide class of QWs, and show that it leads to an entire class of PDEs, encompassing the Hamiltonian form of the massive Dirac equation in $(1+1)$ curved spacetime. Therefore a certain QW, which we make explicit, provides us with a unitary discrete toy model of a test particle in curved spacetime, in spite of the fixed background lattice. Mathematically we have introduced two novel ingredients for taking the continuum limit of a QW, but which apply to any quantum cellular automata: encoding and grouping. 
\end{abstract}

\maketitle

\section{Introduction}

Quantum walks (QW) were originally introduced \cite{BenziSucci, Bialynicki-Birula, MeyerQLGI, Kempe} as dynamics having the following features: \emph{(i)} the underlying spacetime is a discrete grid; \emph{(ii)} the evolution is unitary; \emph{(iii)} it is homogeneous, i.e. translation-invariant and time-independent, and \emph{(iv)} it is causal, i.e. information propagates strictly at a bounded speed. 

Quantum Computing has a number of algorithms that are phrased in terms of QWs, see \cite{venegas2012quantum} for a review. Our focus here is on QWs models {\em per se}, or as models of a given quantum physical phenomena, through a continuum limit. Such QWs models have a broad scope of applications:
\begin{itemize}
  \item they provide quantum algorithms, for the efficient simulation of the modelled phenomena upon a quantum simulation device \cite{FeynmanQC};
  \item even for a classical computer they provide a stable numerical scheme, thereby guaranteeing convergence of the simulation as soon as the scheme is consistent \cite{arrighi2013dirac};
  \item they provide discrete toy models to explore foundational questions \cite{d2014derivation,arrighi2014discrete,ArrighiKG,farrelly2014causal,farrelly2014discrete,LloydQG}.
\end{itemize}

In this paper we introduce \emph{Paired QWs}, which are both a subclass of the general QWs described above, and generalization of the most usual QWs found in the literature.  Basically, \emph{(i)} the input is allowed a simple prior encoding and \emph{(ii)} the local unitary `coin' is allowed to act on larger than usual neighbourhoods. Moreover, the coin is allowed to depend on space and time, as in other QW models. 

We show that Paired QWs admit as continuum limit the class of PDEs of form
\begin{equation}
\partial_t \psi(t,x) = B_1 \partial_x \psi(t,x) + \frac{1}{2} \partial_x B_1 \psi(t,x) + \ii C \psi(t,x) \label{eq:ContLimitQW}
\end{equation}
with $B_1$ and $C$ hermitian and $|B_1|\leq Id$. This class of PDEs includes the Hamiltonian form of the massive curved Dirac equation in $(1+1)$-dimensions \cite{de1962representations} for any bounded metric in any coordinate system, together with an electromagnetic field. Given the PDE we wish to simulate, we are able to retro-engineer the corresponding Paired QW. 

\begin{figure}[t]
\centering
\includegraphics[width=8.5cm,height=4cm]{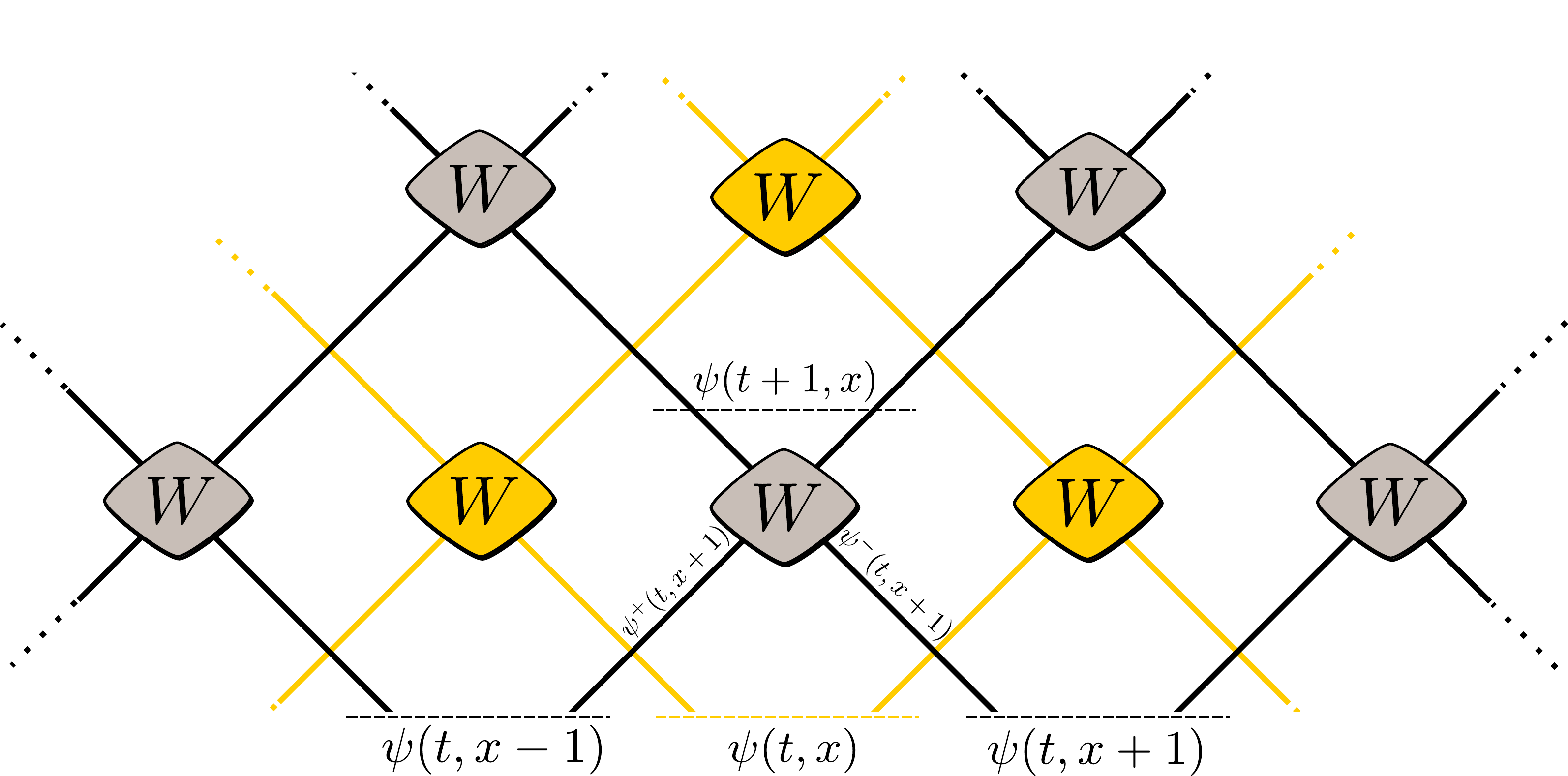} 
\caption{Usual QWs. Times goes upwards. Each site contains a $2d$-dimensional vector $\psi=\psi^+\oplus\psi^-$. Each wire propagates the $d$-dimensional vector $\psi^\pm$. These interact via the $2d\times 2d$ unitary $W$. The circuit repeats infinitely across space and time. Notice that there are two light-like lattices evolving independently.}
\label{fig:LLLat1}
\end{figure}

The results extend the connection between QWs and the Dirac equation, first explored in \cite{BenziSucci,Bialynicki-Birula,MeyerQLGI, bracken2007free}, and further developed in \cite{dariano2012diracca,bisio2013dirac,shikano2013discrete, arrighi2013dirac, farrelly2014discrete, strauch2006relativistic}. Extension to curved spacetime was initiated in \cite{di2013quantum, di2014quantum,succi2015qwalk}, more carefully discussed in the conclusion.

We proceed by first formally defining the model, in Section \ref{sec:Model}. We then compute the conditions for the continuum limit to exist, and provide a complete parametrization of the QW operators in terms of the metric, in Section \ref{sec:ContLimit}.  Then we identify the continuum limit with the Dirac equation in curved spacetime, and validate the model with numerical simulations in Section \ref{sec:DiracMatching}. Finally, we discuss perspectives and related works in Section \ref{sec:Discussion}.

\section{Model definition}\label{sec:Model}

\begin{figure}[t!]
\centering
\includegraphics[scale=0.25]{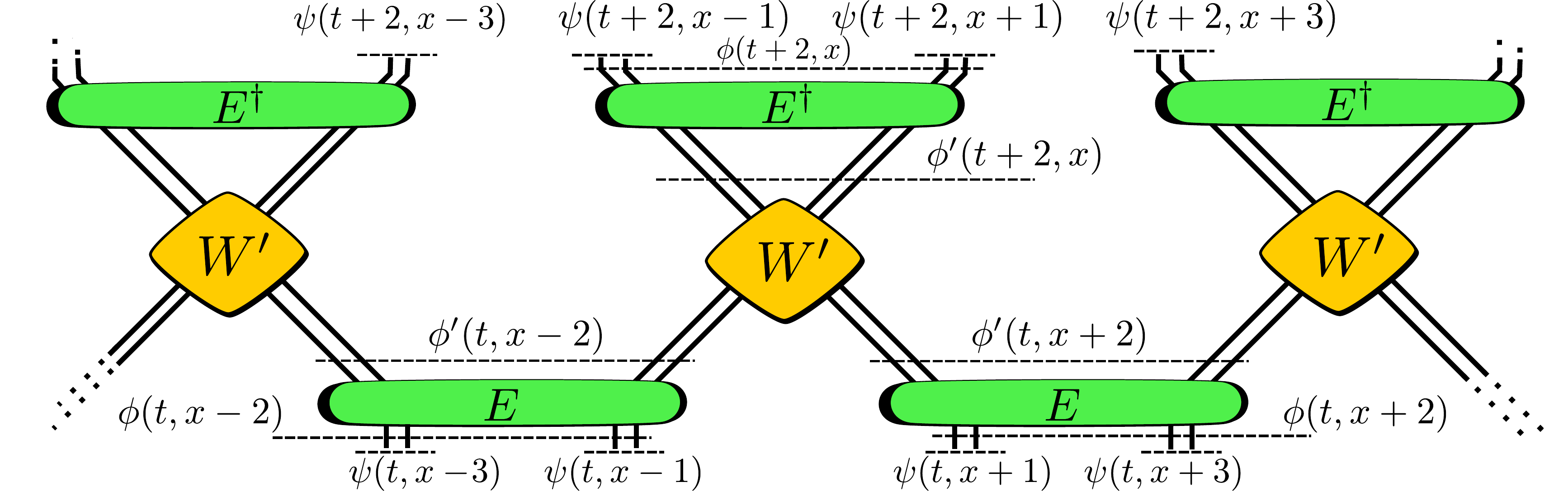} 
\caption{The input to a Paired QW is allowed to be encoded via a unitary $E$, and eventually decoded with $E^\dagger$.}
\label{fig:LLLat2}
\end{figure}

\begin{figure}[t!]
\centering
\includegraphics[width=8.5cm]{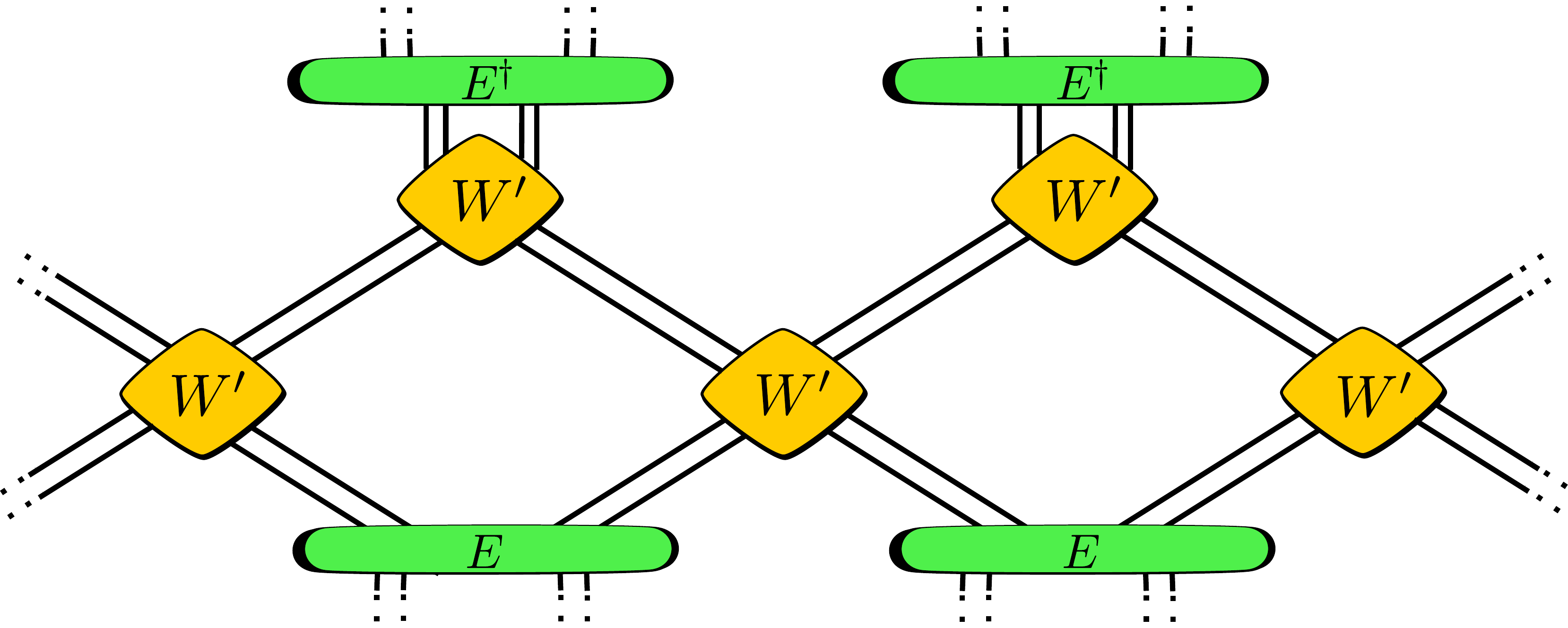} 
\caption{When the scheme is iterated, the decoding of the previous time-step cancels out with the encoding of the next time step. Thus the only relevant encoding/decoding are those of the initial input and final output. A Paired QW is therefore really just a QW, with a particular choice of initial conditions.}
\label{fig:LLLat3}
\end{figure}

Usual one-dimensional QWs act on the space $\ell^2(\Z; \C^{d}\oplus \C^{d})$, equal to the set of square summable sequences in the space $\bigoplus_{\mathbb{Z}} (\C^{d}\oplus\C^{d})$. Often, the dimension of the internal degree of freedom is two, corresponding to $d=1$. We denote $\psi(t)$ those functions taking a lattice position $x$ into the $\C^{2d}$-vector $\psi^+(t,x)\oplus\psi^-(t,x)$, where each $\psi^\pm(t,x)$ is a $\C^{d}$-vector.

These QWs are induced by a local unitary $W$ from $\C^{2d}$ to $\C^{2d}$ often referred to as the coin. Hence $c=2d$ is often referred to as the coin dimension or internal degree of freedom of the walker. The reason why $c$ must split as $d+d$ is because of the way $W$ is wired: each $W(t,x)$ takes one half of $\psi(t,x-1)$ (more precisely, its $d$ upper components $\psi^+(t,x-1)$) and half of $\psi(t,x+1)$ (more precisely, its $d$ lower components $\psi^-(t,x+1)$) in order to produce $\psi(t+1,x)$. This way the inputs and outputs of the different $W(t,x)$ are non-overlapping and they can be applied synchronously  to generate the QW evolution over the full line, so that
\begin{equation}
U(t) := \bigoplus_{x \in \Z} W(t,x)
\end{equation}
generates one time step of the QW (we remark that $t$ indicates possible time dependence of the local unitaries; do not confuse $U(t)$ with the evolution operator from time $0$ to time $t$).

It follows that usual QWs evolve two independent light-like lattices, as emphasized in Fig. \ref{fig:LLLat1}. On one of the light-like lattices, the evolution is given by 
\begin{equation}
V(t) := \bigoplus_{x \in 2\Z} W(t,x)\text{~and~}V(t+1) :=  \bigoplus_{x \in 2\Z + 1} W(t+1,x).
\end{equation}
whilst on the other lattice everything is shifted by $1$ in position,
\begin{equation}
V(t) := \bigoplus_{x \in 2\Z+1} W(t,x)\text{~and~}V(t+1) :=  \bigoplus_{x \in 2\Z } W(t+1,x).
\end{equation}

Paired QWs arise as follows. Bunching up every $\psi(t,x-1)$ and $\psi(t,x+1)$ site into $\phi(t,x)=\psi(t,x-1)\oplus\psi(t,x+1)$, and applying a unitary encoding $E$ to each bunch, we obtain $\phi'(t,x)=E\phi(t,x)$. We may now define a QW over the space $\bigoplus_{2\mathbb{Z}} (\C^{2d}\oplus\C^{2d})$ of these encoded bunches $\phi'$. The local unitary $W'$ will be from $\C^{4d}$ to $\C^{4d}$, and each $W'(t,x)$ will take one half of $\phi'(t,x-2)$ (more precisely, its $2d$ upper components) and half of $\phi(t,x+2)$ (more precisely, its $2d$ lower components) in order to produce $\phi'(t+2,x)$. The inputs and outputs of the different $W'(t,x)$ are again non-overlapping and they can be applied synchronously to generate the QW evolution over the full line, 
\begin{equation}
U(t) := \bigoplus_{x \in 2\Z} W'(t,x).
\end{equation}

In the end, each $\phi'(t+2,x)$ may be decoded as $\phi(t+2,x)=E^\dagger \phi'(t+2,x)$ and be reinterpreted as $\phi(t+2,x)=\psi(t+2,x-1)\oplus\psi(t+2,x+1)$. Clearly this Paired QW (pictured in Figs. \ref{fig:LLLat2} and \ref{fig:LLLat3}) phrased in terms of $\phi'$ and $d'=2d$ is no different from the usual QW definition right above. At least from a discrete point of view.

When looking for a continuum limit, a subtle difference arises. Indeed, say that the regular initial condition is given in terms of the fine-grained spacelike surface of $\psi(t)$, which is assumed to be smooth, i.e. $\psi(t,x)\approx\psi(t,x+1)$. Then the resulting $\phi(t)$ will be smooth both externally, i.e. $\phi(t,x)\approx\phi(t,x+1)$, and internally, i.e. $\phi(t,x)\approx\psi(t,x)\oplus\psi(t,x)$, which is not so usual to ask for. Similarly, $\phi'(t)$ will be smooth both externally, i.e. $\phi'(t,x)\approx\phi'(t,x+1)$ and internally, $\phi'(t,x)\approx E(\psi(t,x)\oplus\psi(t,x))$. It turns out that such reinforced regularity conditions are necessary for some Paired QWs to have a limit.

\begin{figure}[t!]
\centering
\includegraphics[width=\columnwidth]{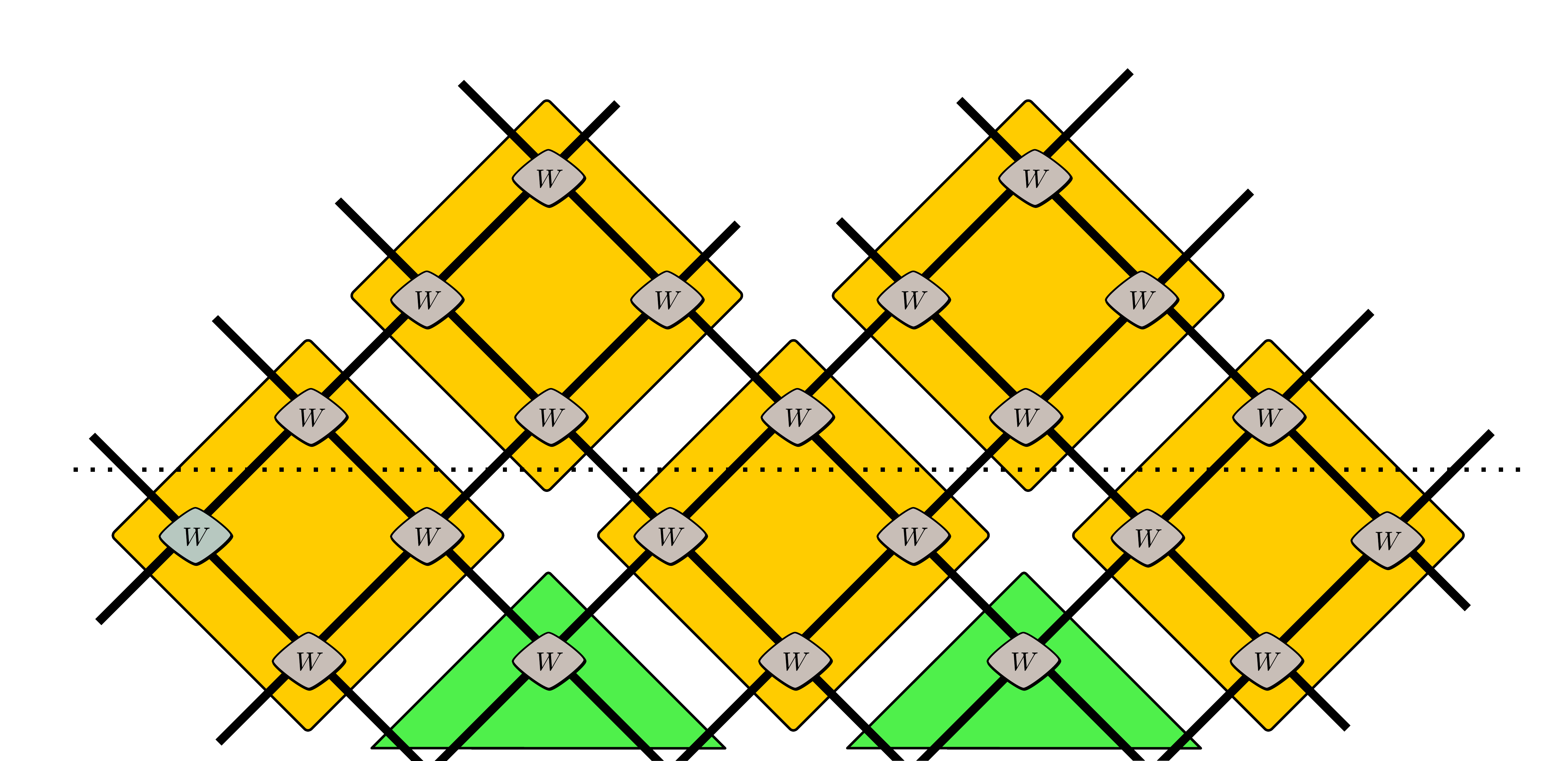} 
\caption{A Paired QW obtained by spacetime grouping of an ordinary QW. The green triangles define the appropriate encoding $E$, that relates the fine-grained input $\psi$ with the coarsegrained input $\phi'$. The dotted line indicates a $t+2$ space-like surface fine-grained output. This surface is recovered by undoing the triangles above these dotted line, which is the role of $E^\dagger$.}
\label{fig:LLLat4}
\end{figure}

The next paragraph is to emphasize that Paired QWs, and their reinforced regularity assumptions, are not ad-hoc: they arise naturally when one performs spacetime grouping of QWs. 

A natural example of Paired QW is provided by performing the spacetime grouping of a usual QW, an operation which we now explain. The spacetime grouping operation takes a QW over $\bigoplus_{\mathbb{Z}} (\C^{d}\oplus\C^{d})$, with local unitary $W$ into a Paired QW over $\bigoplus_{2\mathbb{Z}} (\C^{2d}\oplus\C^{2d})$, with local unitary $W'$, as pictured in Fig. \ref{fig:LLLat4}. 

It is important to notice that if the initial condition was given by $\psi(t)$ for the original walk, the initial condition for the spacetime grouped QW is now given by the $\phi'(t,x)=E(x)\phi(t,x)$, and $\phi(t,x)=\psi(t,x-1)\oplus\psi(t,x+1)$, as pictured in Fig. \ref{fig:LLLat4} again. In the end, each $\phi'(t+2,x)$ may be decoded as $\phi(t+2,x)=E^\dagger(x) \phi'(t+2,x)$ and be reinterpreted as $\phi(t+2,x)=\psi(t+2,x-1)\oplus\psi(t+2,x+1)$. 
This spacetime grouping is reminiscent of the ``stroboscopic'' approach of \cite{di2013quantum,di2014quantum}, but has the advantage of mapping usual QWs into usual QWs of increased dimension. 

This paper studies the continuum limits of Paired QWs (not necessarily arising from a spacetime grouping) for $d=1$, systematically. Recall that for $d=1$: $\psi(t)$ is in $\ell^2(\Z; \C^{2})$ and represents the `physical' field; $\phi'(t)$ is in $\ell^2(\Z; \C^{4})$ and represents a paired, encoded version of it; $W'$ is the $4\times 4$ coin operator.  For our purpose, it will be useful to redefine the bunching-up $\phi(t,x)$ as 
\begin{equation} \label{eq:Groupedwavefunction}
\phi(t,x) := \begin{bmatrix}
u(t,x) \\ d(t,x) \\ u'(t,x) \\ d'(t,x)
\end{bmatrix},
\end{equation}
with

\begin{subequations}\label{eqHad}
\begin{align}
\begin{bmatrix}
u(t,x) \\ u'(t,x)
\end{bmatrix} = H \begin{bmatrix}
\psi^+(t,x+1) \\ \psi^+(t,x-1)
\end{bmatrix} \\ \begin{bmatrix}
d(t,x) \\ d'(t,x)
\end{bmatrix} = H \begin{bmatrix}
\psi^-(t,x+1) \\ \psi^-(t,x-1)
\end{bmatrix}
\end{align}
\end{subequations}
where $H=\frac{1}{\sqrt{2}} \begin{pmatrix}
1 & 1 \\ 1 & -1
\end{pmatrix}$ is the Hadamard matrix.

Notice that the $\psi(t,x)$ dependencies are the same as before, this is just a matter of applying a unitary pre-encoding. This convenient choice of basis is so that in the continuum limit, $u(t,x)$ becomes proportional to $\psi^+(t,x)$, whereas $u'(t,x)$ becomes proportional to the spatial derivative of $\psi^+(t,x)$. 

Armed with those conventions on Paired QW, we can focus on how $\phi_{out} := \phi(t+2,x)$ gets computed, from $\phi_{in} := \phi(t,x-2) \oplus \phi(t,x+2)$. This $\C^{4}\oplus\C^{4}$ to $\C^{4}$ function may be thought of as the local rule of a cellular automata with cells in $\C^{4}$. Its explicit formula is given by 
\begin{align}
G &= E^\dagger(t+2,x) W'(t,x) (P' \oplus P) \nonumber \\ & \qquad (E(t,x-2)\oplus E(t,x+2)), \label{eq:FormalDefGQW}
\end{align}
where the $2\times 4$ projectors $P$ and $P'$ pick-up the $u,d$ (non-primed subspace) and $u',d'$ (primed subspace) coordinates, respectively. Thus
\begin{equation}
\phi(t+2,x) = G (\phi(t,x-2) \oplus \phi(t,x+2)). \label{eq:FormalGQWEq}
\end{equation}

\section{Continuum limit} \label{sec:ContLimit}

From now on, we consider that $t$ and $x$ are continuous variables, and choose the same discretization step $\varepsilon \in \R^+$ for each coordinate. In particular note that, to first order in $\varepsilon$, we have that $u \simeq \sqrt{2} \psi^+$, $d \simeq \sqrt{2} \psi^-$, $u' \simeq \varepsilon \sqrt{2} \partial_x \psi^+$ and $d' \simeq \varepsilon \sqrt{2} \partial_x \psi^-$, see \eqref{eqHad}. To start investigating the continuum limit of the system defined by Eq. \eqref{eq:FormalDefGQW}, we compute the expressions for the input and output.

The expansion of the input to first order in $\varepsilon$ in terms of $u,u',d,d'$ is
\begin{equation}
\phi_{in}(t,x) \simeq \begin{bmatrix} u \\ d \\ 0 \\ 0 \end{bmatrix} \oplus \begin{bmatrix} u \\ d \\ 0 \\ 0\end{bmatrix} +
\begin{bmatrix} -2u' \\ -2d' \\ u' \\ d' \end{bmatrix} \oplus \begin{bmatrix} 2u' \\ 2d' \\ u' \\ d' \end{bmatrix}. \label{eq:phiinDev}
\end{equation}
We stress that $u'$ and $d'$ are themselves proportional to $\varepsilon$, hence the last term is proportional to $\varepsilon$.

The expansion of the output to first order in $\varepsilon$ in terms of $u,u',d,d'$ is
\begin{equation}
\phi_{out}(t,x)  \simeq
  \begin{bmatrix} u \\ d \\ 0 \\ 0 \end{bmatrix} +
  \begin{bmatrix} 2\varepsilon \partial_t u \\ 2\varepsilon\partial_t d \\ u' \\ d' \end{bmatrix}. \label{eq:phioutDev}
\end{equation}

Next we specify the structure of the walk and encoding operators. We shall assume, for simplicity, that the matrix elements of $W$ and $E$ are analytic functions of $(t,x)$ and $\varepsilon$. 

First, we set $W' := W^{(0)}e^{\ii \varepsilon \tilde{W}}$, with $W^{(0)}$ unitary and $\tilde{W}$ hermitian. This enforces the unitarity of $W'$, and is without loss of generality, since only its expansion to first order in $\varepsilon$ matters: 
\begin{equation}
W(t,x) \simeq W^{(0)}(t,x) + \ii \varepsilon W^{(0)}(t,x)\tilde{W}(t,x). \label{eq:Wexpansion}
\end{equation}

Then, in a similar manner, we define $E := E^{(0)}e^{\ii \varepsilon \tilde{E}}$, with $E^{(0)}$ unitary and $\tilde{E}$ hermitian. Hence, to first order in $\varepsilon$,
\begin{equation}
E(t,x) \simeq E^{(0)}(t,x) + \ii\varepsilon E^{(0)}(t,x)\tilde{E}(t,x). \label{eq:Eexpansion}
\end{equation}

Here is some notation. Any matrix $A \in \C^{4\times 4}$ will be written in block form as $A = \begin{pmatrix}
A_1 & A_3 \\ A_2 & A_4
\end{pmatrix}$, where $A_j \in \C^{2\times 2}$, $j=1,\ldots,4$. Let $X=\sigma_x\otimes I$, $Y=\sigma_y \otimes I$ and $Z = \sigma_z \otimes I$, where $(\sigma_x,\sigma_y,\sigma_z)$ are the Pauli spin matrices. 

Notice that, for any $A \in \C^{4\times 4}$, the following simplifications hold:
\begin{equation}
(P' \oplus P)(A\oplus A)(v\oplus v) = X A v \qquad \forall v\in \C^4 \label{eq:Simplif1}
\end{equation}
and
\begin{equation}
(P' \oplus P)(A\oplus A)(-v\oplus v) =  X Z A v \qquad \forall v\in \C^4. \label{eq:Simplif2}
\end{equation}

Next we develop the zeroth order and the first order expansion in $\varepsilon$ of Eq. \eqref{eq:FormalDefGQW}. 

\subsection{Zeroth order} 

For the left hand side we have just the zeroth order of \eqref{eq:phioutDev}, while for the right hand side there is only one term which does not contain $\varepsilon$, obtained multiplying all the zeroth order contributions. Hence 
\begin{align}
\begin{bmatrix} u \\ d \\ 0 \\ 0 \end{bmatrix} &= E^{(0)\dagger} W^{(0)} (P' \oplus P)(E^{(0)}\oplus E^{(0)})
\begin{bmatrix} u \\ d \\ 0 \\ 0 \end{bmatrix} \oplus \begin{bmatrix} u \\ d \\ 0 \\ 0\end{bmatrix} \nonumber \\
&=  E^{(0)\dagger} W^{(0)} X E^{(0)} \begin{bmatrix} u \\ d \\ 0 \\ 0 \end{bmatrix},  \label{eq:ZerothOrder}
\end{align}
where we used the simplification \eqref{eq:Simplif1}. The only non-trivial relations are 
\begin{align}
\begin{bmatrix}
u \\ d 
\end{bmatrix} &= \left( E^{(0)\dagger} W^{(0)} X E^{(0)} \right)_1 \begin{bmatrix}
u \\ d 
\end{bmatrix} \label{eq:condZerothOrderA} \\
\begin{bmatrix}
0 \\ 0 
\end{bmatrix} &= \left( E^{(0)\dagger} W^{(0)} X E^{(0)} \right)_2 \begin{bmatrix}
u \\ d 
\end{bmatrix} \label{eq:condZerothOrderB}
\end{align}
To satisfy \eqref{eq:condZerothOrderA} for arbitrary $u$ and $d$, we must take the identity for block $1$. Now, since the matrix in \eqref{eq:ZerothOrder} is unitary, then both its rows and its columns must sum to one, thus the blocks $2$ and $3$ become zero, and \eqref{eq:condZerothOrderB} is automatically satisfied; we are left with the choice of an arbitrary unitary $U \in U(2)$ for block $4$, to complete the matrix. Hence  
\begin{equation}
  E^{(0)\dagger} W^{(0)} X E^{(0)} = I \oplus U, \label{eq:ZerothOrderC}
\end{equation}
where the direct sum is respect to the non-primed subspace (spanned by the first two entries) and the primed subspace (spanned by the last two entries).

\subsection{First order} 

For the left hand side we have just the first order of \eqref{eq:phioutDev}; note it contains time and space derivatives of $\psi^\pm$. For the right hand side we multiply and collect all possible combinations in which only one term contains $\varepsilon$. Then, after a long but straightforward calculation (see Appendix \ref{app:CalculationFirstOrder} for the details), we get
\begin{align}
\begin{bmatrix}2 \varepsilon \partial_t u \\ 2\varepsilon \partial_t d \\ u' \\ d'\end{bmatrix} &=  (I \oplus U) \begin{bmatrix} 0 \\ 0 \\ u' \\ d' \end{bmatrix} + (I \oplus U) B \begin{bmatrix} 2u' \\ 2d' \\ 0 \\ 0 \end{bmatrix} \nonumber \\ &+ \varepsilon \left\{ (2N -\ii\tilde{E})(I\oplus U) \right. \nonumber \\ &\left.+  (I \oplus U) (\ii\tilde{E}+2M) + T   \right\} \begin{bmatrix}
    u \\ d \\ 0 \\ 0
  \end{bmatrix}. \label{eq:ContLimitFirstOrder}
\end{align}
with
\begin{subequations}
\begin{align}
B &= E^{(0)\dagger} Z E^{(0)} \label{eq:BigB}\\
N &= (\partial_t E^{(0)\dagger}) E^{(0)} \label{eq:BigN} \\
T &= \ii E^{(0)\dagger} W^{(0)} \tilde{W} X E^{(0)}\label{eq:BigT}\\
M &= E^{(0)\dagger} Z (\partial_x E^{(0)})  \label{eq:BigM}.
\end{align}
\end{subequations}
To deal with \eqref{eq:ContLimitFirstOrder} we shall study separately what happens in the primed and in the non-primed subspaces.

\subsection{Continuum limit equation} 

Projecting Eq. \eqref{eq:ContLimitFirstOrder} on the non-primed subspace, we obtain an equation with time derivatives in the left hand side, 
\begin{equation}
\begin{bmatrix}
2\varepsilon \partial_t u \\ 2\varepsilon \partial_t d
\end{bmatrix} = B_1 \begin{bmatrix}
2u' \\ 2d'
\end{bmatrix} + \varepsilon ( 2 N_1 + T_1 + 2M_1  ) \begin{bmatrix}
u \\ d
\end{bmatrix}.
\end{equation}
Switching to the original $\psi^\pm(t,x)$ coordinates, and writing $\psi(t,x)=[\psi^+(t,x),\psi^-(t,x)]^\transp$, 
\begin{equation}
\partial_t \psi(t,x) = B_1 \partial_x \psi(t,x) + \left(N_1 + \frac{T_1}{2}+M_1 \right) \psi(t,x).
\end{equation}

From \eqref{eq:BigB}, applying Leibniz rule and using \eqref{eq:BigM} we have\footnote{Recall that if $A \in \C^{n\times n}$, its real and imaginary parts are $\Re A := \frac{1}{2}( A+A^\dagger)$ and $\Im A := \frac{1}{2\ii} (A-A^\dagger)$, respectively.}
\begin{equation}
\partial_x B = M + M^\dagger = 2\Re M. \label{eq:ReldXBM}
\end{equation}

From \eqref{eq:BigN}, the unitarity of $E^{(0)}$ implies that $N$ is skew-hermitian,
\begin{equation}
N^\dagger = -N. \label{eq:RelNSH}
\end{equation}

From \eqref{eq:BigT},
\begin{align}
T &= \ii E^{(0) \dagger} W^{(0)} \tilde{W} X E^{(0)} \\ &= \ii E^{(0) \dagger} W^{(0)} X E^{(0)} E^{(0) \dagger} X \tilde{W} X E^{(0)}  \\ &= \ii (I\oplus U) E^{(0) \dagger} X \tilde{W} X E^{(0)},
\end{align}
where we used the zeroth order condition \eqref{eq:ZerothOrderC}. Inverting,
\begin{align}
\ii E^{(0) \dagger} X \tilde{W} X E^{(0)} = (I\oplus U^\dagger)T =
\begin{pmatrix}
T_1 & T_3 \\ U^\dagger T_2 & U^\dagger T_4
\end{pmatrix}. \label{eq:tildeWrelT}
\end{align}
Since the left hand term is skew-hermitian we have that
\begin{subequations}
\begin{align}
T_1^\dagger &= -T_1  \label{eq:CondT1SH}\\
T_3 &= -T_2^\dagger U \label{eq:CondT3SH} \\
T_4^\dagger U &= -U^\dagger T_4. \label{eq:CondT4SH}
\end{align}
\end{subequations}

Therefore, by spliting $M_1$ into its hermitian and skew-hermitian parts, and using equations \eqref{eq:ReldXBM}, \eqref{eq:RelNSH} and \eqref{eq:CondT1SH}, the continuum limit has the general form
\begin{equation}
\partial_t \psi(t,x) = B_1 \partial_x \psi(t,x) + \frac{1}{2} \partial_x B_1 \psi(t,x) + \ii C \psi(t,x). 
\end{equation}
where $C$ is an hermitian matrix defined by
\begin{equation}
\ii C = N_1 + \frac{T_1}{2} + \ii\Im M_1. \label{eq:defC}
\end{equation}

\subsection{Compatibility constraints} 

Projecting Eq. \eqref{eq:ContLimitFirstOrder} onto the primed subspace, gives
\begin{align}
\begin{bmatrix}
u' \\ d'
\end{bmatrix} &= U \begin{bmatrix}
u' \\ d'
\end{bmatrix} + 2U B_2 \begin{bmatrix}
u' \\ d'
\end{bmatrix} + \varepsilon \left( 2N_2 -\ii \tilde{E}_2  \right. \nonumber \\ &\qquad + \left. \ii U\tilde{E}_2 + 2UM_2 + T_2 \right) \begin{bmatrix}
u \\ d
\end{bmatrix}. \label{eq:CompatConstraint}
\end{align}

Eq. \eqref{eq:CompatConstraint} does not involve time derivatives. Therefore, these equations must be understood as constraints. Where do these come from? Recall that the aimed continuum limit equation \eqref{eq:ContLimitQW} is over a $\mathbb{C}^2$ field, but the QW employed is over the $\mathbb{C}^4$ field obtained by pairing it. Thus, the $\mathbb{C}^4$ field has some internal smoothness initially, which the QW must preserve. More precisely, in order to have nontrivial, time-dependent solutions, the coeffients of $[u, v]^\transp$ and $[u', v']^\transp$ must vanish separately:
\begin{subnumcases}{}
U(I+2 B_2) = I, \label{eq:condUB} \\
2N_2  -\ii (I-U) \tilde{E}_2 + 2UM_2 + T_2 = 0. \label{eq:condNMT}
\end{subnumcases}

\subsection{Existence of solutions}

Up to now we have determined the continuum limit, provided that the constraints \eqref{eq:condUB}-\eqref{eq:condNMT} are satisfied. In this section we show that, given any hermitian $B_1$ and $C$, there are indeed compatible choices of $W$ and $E$.

The strategy is the following. First we show that $B_1$ along with constraint \eqref{eq:condUB} determines the zeroth order part of $E$ and $W'$.
Then, using $C$ and \eqref{eq:condNMT} we complete the solution by determining the first order terms.

\subsubsection{Determination of $B$ and $U$}

Consider the spectral decomposition $B_1 = VDV^\dagger$, $D = \diag \{ d_1,d_2\}$. In Appendix \ref{app:GeneralB} we show that Eq. \eqref{eq:BigB} implies that $d_1,d_2$ must belong to the interval $[-1,1]$, and provide the general form of $B$ given $B_1$ (see section \ref{sec:DiracMatching} for a discussion about the eigenvalue constraint). Here we just pick one particular solution, namely
\begin{equation}
B = \begin{pmatrix}
V^\dagger & 0 \\ 0 & V^\dagger
\end{pmatrix} \overline{B} \begin{pmatrix}
V & 0 \\ 0 & V
\end{pmatrix}, \label{eq:OvBV}
\end{equation}
where $\overline{B}$ is
\begin{equation}
\overline{B} = \begin{pmatrix}
d_1 & 0 & -\lambda_1 e^{\ii \eta_1} & 0 \\
0 & d_2 & 0 & -\lambda_2 e^{\ii \eta_2} \\
-\lambda_1 e^{-\ii \eta_1} & 0 & -d_1 & 0 \\
0 & -\lambda_2 e^{-\ii \eta_2} & 0 & -d_2  
\end{pmatrix}, \label{eq:Bbar}
\end{equation}
with $\lambda_i = \sqrt{\smash[b]{1-d_i^2}}$, $\sin \eta_i = \pm |d_i|$, $-\pi/2 < \eta_i < \pi/2$, $i \in \{1,2\}$. 

Note that $U$ is now fixed by Eq. \eqref{eq:condUB}.

\subsubsection{Determination of $E^{(0)}$ and $W^{(0)}$}

From Eq. \eqref{eq:BigB} we know that $E^{\dagger (0)}$ diagonalizes $B$. Then, its columns can be chosen to be any complete set of normalized eigenvectors of $B$. More generally, we could take $\begin{pmatrix} R & 0 \\ 0 & S \end{pmatrix} E^{(0)}$ for arbitrary $R, S \in U(2)$, because of the degeneracy of order two for each eigenvalue $+1$, $-1$.

For the special case of $\overline{B}$ in Eq. \eqref{eq:Bbar}, we can give an explicit solution $\overline{E}^{(0)}$, 
\begin{equation}
\overline{E}^{(0)} = \frac{1}{\sqrt{2}} \begin{pmatrix}
\nu_1^+ & 0 & - \nu_1^- e^{\ii \eta_1} & 0 \\
0 & \nu^+_2 & 0 & -\nu_2^- e^{\ii \eta_2} \\
\nu_1^-  & 0 & \nu_1^+ e^{\ii \eta_1} & 0 \\
0 & \nu_2^- & 0 & \nu_2^+ e^{\ii \eta_2}  
\end{pmatrix}, \label{eq:EBar0}
\end{equation}
where $\nu_i^\pm = \sqrt{1\pm d_i}$, $i\in \{1,2\}$. 

Once $E^{(0)}$ is known, we can compute $W^{(0)}$ from \eqref{eq:ZerothOrderC}.

\subsubsection{Determination of $\tilde{E}$ and $\tilde{W}$}

The choice of $E^{(0)}$ determines $N_1$ and $M_1$ via Eqs. \eqref{eq:BigN} and \eqref{eq:BigM}, so $T_1$ is fixed by Eq. \eqref{eq:defC} once we choose $C$.

Since $\tilde{E}$ does not appear in the continuum limit, without loss of generality we can take $\tilde{E} = 0$. In this way $T_2$ is fixed by the contraint \eqref{eq:condNMT}.

In order to complete $T$ it is now sufficient to take $T_4=0$, and $T_3$ from \eqref{eq:CondT3SH}. Finally, from \eqref{eq:tildeWrelT} we find $\tilde{W}$,
\begin{equation}
\tilde{W} = -\ii X E^{(0)} (I\oplus U^\dagger) T E^{(0) \dagger} X. \label{eq:DefWtilde}
\end{equation}

\subsection{Recap} 

We have shown that the continuum limit of our model is given by Eq. \eqref{eq:ContLimitQW}. Moreover, we provided a procedure to obtain the parameters of the quantum walk, namely the unitaries $W'$ and $E$, given a pair of hermitian matrices $B_1$ and $C$, possibly spacetime dependent. We remark that the choices made in the procedure are in general not unique. In the emergent continuum limit, different choices of $E$, $W'$ lead in general to the same equation. 

Notice also that the minimal coupling (e.g. electric field) is already considered in the parameter $C$.

The whole procedure was programmed in {\tt sagemath}, and made available in \cite{onlinesage}.

\begin{figure}[t]
\centering
\includegraphics[width=\columnwidth]{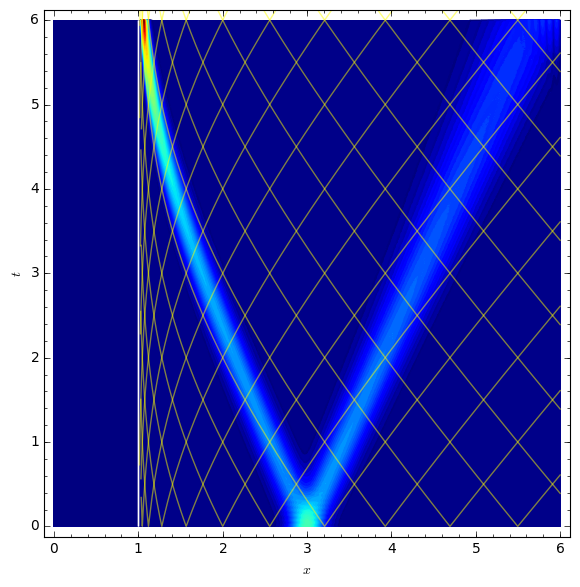} 
\caption{Simulation of the Paired QW for the Schwarzschild metric with mass parameter $M=0.5$. We plot the probability density for a particle with initial condition given by a gaussian wavepacket $\phi(x) \propto \int e^{-(p-p_0)^2/(2\sigma^2)+\ii (x-x_0)p} \left(u_+(p) + u_-(p)\right) dp$ where $x_0=3.0$, $p_0=50$, $\sigma=1.56$ and $u_\pm$ are the eigenvectors of the free Dirac Hamiltonian $H_0 = \alpha \hat{p} + m \beta$. The mass of the particle is $m=50$. For comparison, we show in yellow a grid of null geodesics. The lattice spacing is $\varepsilon = 5 \times 10^{-5}$.}
\label{fig:SCHMETRIC1}
\end{figure}

\section{Recovering the Dirac equation} \label{sec:DiracMatching}

On a spacetime with metric tensor $g_{\mu\nu}$ and in the absence of external fields, the Dirac equation in Hamiltonian form \cite{de1962representations} is $\ii \partial_t \psi = H_D \psi$, with
\begin{equation}
H_D = -\ii\left(\alpha\frac{e^1_1}{e^0_0} + e^1_0\right)\partial_x - \frac{\ii}{2} \partial_x\left(\alpha\frac{e^1_1}{e^0_0} + e^1_0\right) + \frac{m}{e^0_0} \beta, \nonumber 
\end{equation}
where $m$ is the mass and $\alpha, \beta$ are matrices satisfying $\alpha^2=\beta^2=I$, $\alpha\beta+\beta\alpha=0$ (here and in the following we assume natural units, $\hbar = c = G = 1$). Here $e^{\mu}_a(t,x)$ are the \emph{dyads}, which are related to the metric via $g_{\mu\nu}(t,x)e^\mu_a(t,x) e^\nu_b(t,x)  = \eta_{ab}$, where $\eta_{ab}$ is the Minkowski metric.

Making the identification with the continuum limit of our discrete model, Eq. \eqref{eq:ContLimitQW}, we find
\begin{align}
  B_1 &= -\frac{e^1_1}{e^0_0}\alpha - e^1_0 \\
  C &= - \frac{m}{e^0_0} \beta.
\end{align}
These equations allow to find the QW parameters, associated to a given metric. The constraint that the eigenvalues of $B_1$ are $d_{1,2} \in [-1,1]$ represents the finite speed of propagation on the lattice. In practice, for any region of spacetime where the metric field is bounded, it is possible to rescale the coordinates in such a way that the physical lightcones are inside the ``causal lightcones'' of the discrete model.

For instance, we can specialize the previous considerations to the case of the Schwarzschild metric, whose radial part is
\begin{equation}
  ds^2 = (1-2M/x)dt^2 - (1-2M/x)^{-1}dx^2,
\end{equation}
where $x$ corresponds to the radial coordinate \footnote{In this case the dyads are $e^0_0=(1-2M/x)^{-1/2}$, $e^1_1=(1-2M/x)^{1/2}$, and $e^0_1=e^1_0=0$.}. Choosing the chiral representation \cite{Thaller}, namely $\alpha =\sigma_z$ and $\beta=\sigma_x$, we have
\begin{align}
B_1 & = -\left(1-\frac{2M}{x}\right)\sigma_z \\
C &= -m\left(1-\frac{2M}{x} \right)^{1/2}\sigma_x.
\end{align}
A simulation for a particle in the Schwarzschild metric is shown in Fig. \ref{fig:SCHMETRIC1}. Again the {\tt sagemath} program which converts any metric into the corresponding QW, and produces such simulations, is available at \cite{onlinesage}.

\section{Summary, related and future works} \label{sec:Discussion}

In summary, we have constructed a QW, i.e. a strictly unitary, causal, local evolution, which implements the idea of ``encoding-evolution-decoding'' of the discrete dynamics, defined in Eq. \eqref{eq:FormalGQWEq}. We have found that Paired QWs with $d=1$ admit as continuum limit all PDEs of the form \eqref{eq:ContLimitQW}. This class encompasses the Hamiltonian form of the Dirac equation in curved spacetime, together with an electromagnetic field.

In this model, curvature is effectively implemented by a collection of spacetime dependent local unitaries (namely, the unitary operators $E(t,x)$ and $W'(t,x)$), which are distributed over a fixed background lattice, and whose purpose is to drive the particle according to the metric. 

Our main contribution in this paper is to extend the existent literature in two directions: we allow for massive particles and for any bounded metric field over an arbitrary coordinate system. Let us explain the relation between the present paper and related works \cite{di2013quantum, di2014quantum} in more detail. In \cite{di2013quantum, di2014quantum}, Di Molfetta et al. have systematically studied the continuum limit of a two-time-step QW with arbitrary coin. This stroboscopic approach made it possible to recover the $(1+1)$-Weyl equation in curved spacetime, for metrics of having $g_{00}=1$. Such metrics exclude the Schwarzschild metric of Fig. \ref{fig:SCHMETRIC1}, for instance. Yet, any metric can be brought to have $g_{00}=1$ if one is allowed a prior change of coordinates. On the other hand, in $(1+1)$-dimensions, every space-time is conformally flat up to a change of coordinates, and therefore allowing changes of coordinates could be said to oversimplify the problem (for conformally flat metrics the Dirac equation can be obtained from a usual QW with a spacetime dependent coin \cite{unplublishedQWMassive}) --- besides breaking away from general covariance. Yet, \cite{di2013quantum, di2014quantum} was no doubt an inspiration : their model can be recovered a specific instance of the Paired QW described in this paper, by a spacetime grouping as described in Fig. \ref{fig:LLLat4}.

Another approach, recently pursued by Succi et al. \cite{succi2015qwalk}, is within the framework of lattice discretization of the relativistic quantum wave equation (quantum lattice boltzmann \cite{succi2001lattice}). The key observation is that the mass term can be recovered by extending the neighbourhood of the dynamical map. However, the existence of a parametrization for the unitary evolution in terms of the metric, has remained an open question, which is solved by this analysis.

Surprisingly, the slight overgenerality of Eq. \eqref{eq:ContLimitQW} with respect to the $(1+1)$ curved Dirac equation, just matches some terms arising as $(1+1)$ projections of the $(2+1)$ curved Dirac equation. This suggests a possible generalization to $(2+1)$ dimensions, through operator splitting, which is the subject of current work. The extension to $(3+1)$ dimensions remains another interesting question, since gamma matrices of the Dirac equation then become four dimensional. In the terminology of this paper this means having to deal with Paired QWs having $d=2$, so that the internal degrees of freedom of the field is four-dimensional. Most of our techniques should carry through, except for the characterization of the family of solutions to the constraints. Another challenging problem is the study the underlying symmetries of the discrete model, e.g. by making explicit some form of discrete general covariance along the same lines as \cite{arrighi2014discrete}.

\begin{acknowledgments}
This work has been funded by the ANR-12-BS02-007-01 TARMAC grant, the ANR-10-JCJC-0208 CausaQ grant, and the John Templeton Foundation, grant ID 15619. The authors acknowledge helpful discussions with Giacomo D'Ariano and Fabrice Debbasch.
\end{acknowledgments}

\begin{widetext}

\appendix

\section{Calculation of the first order expansion of the discrete model} \label{app:CalculationFirstOrder}

In this section we prove Eq. \eqref{eq:ContLimitFirstOrder}, which we reproduce here: 
\begin{align}
\begin{bmatrix}2 \varepsilon \partial_t u \\ 2\varepsilon \partial_t d \\ u' \\ d'\end{bmatrix} &=  (I \oplus U) \begin{bmatrix} 0 \\ 0 \\ u' \\ d' \end{bmatrix} + (I \oplus U) B \begin{bmatrix} 2u' \\ 2d' \\ 0 \\ 0 \end{bmatrix} + \varepsilon \left\{ (2N-\ii\tilde{E})(I\oplus U) \right. \left.+  (I \oplus U) (\ii\tilde{E}+2M) + T   \right\} \begin{bmatrix}
    u \\ d \\ 0 \\ 0
  \end{bmatrix}. 
\end{align}

Recall that we want to expand 
\begin{align}
\phi_{out}(t,x) & = G~\phi_{in}(t,x), \label{eq:EvApp}
\end{align}
where
\begin{align}
G &= E^\dagger(t+2,x) W'(t,x) (P' \oplus P) (E(t,x-2)\oplus E(t,x+2)). \label{eq:EvApp2}
\end{align}
The first order expansion of the encoding and of the walk is, by definition,
\begin{align}
E(t,x) &= E^{(0)}(t,x) + \varepsilon \ii E^{(0)}(t,x)\tilde{E}(t,x) + O(\varepsilon^2) \\
W'(t,x) &= W^{(0)}(t,x) + \varepsilon \ii W^{(0)}(t,x)\tilde{W}(t,x) + O(\varepsilon^2),
\end{align}
hence, to first order in $\varepsilon$, the operators in \eqref{eq:EvApp2} expand to
\begin{align}
&E^{\dagger}(t+2,x) \simeq E^{(0)\dagger} + \varepsilon \left( 2 \partial_t E^{(0)\dagger} -\ii \tilde{E}E^{(0)\dagger} \right)  \\
&W'(t,x) \simeq W^{(0)}+\varepsilon \ii W^{(0)}\tilde{W}\\
&E(t,x-2)\oplus E(t,x+2) \simeq \left( E^{(0)} - 2 \varepsilon \partial_x E^{(0)} + \ii\varepsilon E^{(0)}\tilde{E} \right) \oplus   \left( E^{(0)} + 2 \varepsilon \partial_x E^{(0)} + \ii\varepsilon E^{(0)}\tilde{E} \right) \\
\end{align}
where in the right hand side all operators are evaluated at $(t,x)$. Recall that the first order expansions of the output and input are 
\begin{equation}
\phi_{out}(t,x)  \simeq
  \begin{bmatrix} u \\ d \\ 0 \\ 0 \end{bmatrix} +
  \begin{bmatrix} 2\varepsilon \partial_t u \\ 2\varepsilon\partial_t d \\ u' \\ d' \end{bmatrix}, \qquad 
\phi_{in}(t,x) \simeq \begin{bmatrix} u \\ d \\ 0 \\ 0 \end{bmatrix} \oplus \begin{bmatrix} u \\ d \\ 0 \\ 0\end{bmatrix} +
\begin{bmatrix} -2u' \\ -2d' \\ u' \\ d' \end{bmatrix} \oplus \begin{bmatrix} 2u' \\ 2d' \\ u' \\ d' \end{bmatrix}.
\end{equation}

We shall use the identities
\begin{align}
&( P' \oplus P) (E \oplus E) (v \oplus v) = XEv \\
& (P' \oplus P) (E \oplus E) (-v \oplus v) = XZEv 
\end{align}
valid for any $v \in \C^4$, because $P'$ (resp. $P$) are the projections onto the primed (resp. non-primed) coordinates; in matrix form, 
\begin{equation}
P' = \begin{pmatrix}
0 & 0 & 1 & 0 \\ 0 & 0 & 0 & 1
\end{pmatrix}, \qquad P = \begin{pmatrix}
1 & 0 & 0 & 0 \\ 0 & 1 & 0 & 0
\end{pmatrix}.
\end{equation}

Next we plug the previous expansions into \eqref{eq:EvApp}. Collecting all the terms of first order in $\varepsilon$,
\begin{align*}
\begin{bmatrix} 2\varepsilon \partial_t u \\ 2\varepsilon\partial_t d \\ u' \\ d' \end{bmatrix} &=
 E^{(0) \dagger} W^{(0)} X E^{(0)} \begin{bmatrix} 0 \\ 0 \\ u' \\ d' \end{bmatrix} + E^{(0) \dagger} W^{(0)} X Z E^{(0)} \begin{bmatrix} 2u' \\ 2d' \\ 0 \\ 0 \end{bmatrix} \\ &+\left\{ \varepsilon (2\partial_t E^{(0)\dagger } -\ii \tilde{E}E^{(0) \dagger} )W^{(0)} X E^{(0)} + \ii \varepsilon E^{(0)\dagger} W^{(0)} \tilde{W} X E^{(0)} + \ii \varepsilon  E^{(0)\dagger}W^{(0)} X E^{(0)}\tilde{E} \right. \\ & +\left. 2 \varepsilon E^{(0)\dagger } W^{(0)} X Z \partial_x E^{(0)} \right\} \begin{bmatrix} u \\ d \\ 0 \\ 0 \end{bmatrix}.
\end{align*}

Next we use the zeroth order condition (cf. \eqref{eq:ZerothOrderC}), namely $ E^{(0)\dagger} W^{(0)} X E^{(0)} = I \oplus U$, so that
\begin{align*}
\begin{bmatrix} 2\varepsilon \partial_t u \\ 2\varepsilon\partial_t d \\ u' \\ d' \end{bmatrix} &= (I \oplus U)  \begin{bmatrix} 0 \\ 0 \\ u' \\ d' \end{bmatrix} + (I\oplus U) \underbrace{E^{(0) \dagger} Z E^{(0)}}_{\mathrm{B }}  \begin{bmatrix} 2u' \\ 2d'\\ 0 \\ 0 \end{bmatrix} \\ &\varepsilon \left\{ \left[ 2\underbrace{(\partial_t E^{(0) \dagger})E^{(0)}}_{\mathrm{N}} -\ii\tilde{E} \right] (I\oplus U ) + \underbrace{\ii E^{(0)\dagger} W^{(0)}\tilde{W} X E^{(0)}}_{\mathrm{T}} \right. \\ & \left. + \ii (I\oplus U) \tilde{E} + 2 (I\oplus U) \underbrace{E^{(0)\dagger} Z \partial_x E^{(0)}}_{\mathrm{M}} \right\}  \begin{bmatrix} u \\ d \\ 0 \\ 0 \end{bmatrix},
\end{align*}
and we get the desired result.

\section{General form of $B$} \label{app:GeneralB} 

 Since $B$ must be hermitian, cf. \eqref{eq:BigB}, then $B_1$ and $B_4$ are hermitian. Since it is also unitary, then it must square to the identity. This implies that the conditions 
\begin{subequations}
\begin{align}
B_1^2 + B_2^\dagger B_2 &= \Id_2  \label{eq:condBDiagonal} \\ 
B_4^2 + B_2 B_2^\dagger &= \Id_2 \label{eq:condBDiagonal2}
\end{align}
\end{subequations}
and
\begin{subequations}
\begin{align}
B_2 B_1 + B_4 B_2 &= 0  \label{eq:condBAntiDiagonal} \\ 
B_1 B_2^\dagger + B_2^\dagger B_4 &= 0   \label{eq:condBAntiDiagonal2} 
\end{align}
\end{subequations}
must hold. Note also that $B$ must have a complete set of orthonormal eigenvectors, eigenvalues $\pm 1$, and it shall be traceless, because it is is similar to $Z$. 

First, we parametrize the block $B_2$. Consider the spectral decomposition of $B_1 = V D V^\dagger$, $D = \diag \{ d_1,d_2\}$. From the first of conditions \eqref{eq:condBDiagonal}, we have that $d_1,d_2 \in [-1,1]$, because the square root of the components of $\Id-D^2$ are precisely the singular values of $B_2$, which should be non-negative. Next, we shall find $B_2$ such that constraint \eqref{eq:condUB} is satisfied. The same equation also determines $U$.

We look for $B_2 \in \C^{2\times 2}$ such that conditions \eqref{eq:condUB} and \eqref{eq:condBDiagonal} are satisfied, namely that
\begin{enumerate}
\item $\Id + 2 B_2$ is unitary,
\item $B_2^\dagger B_2 = \Id - B_1^2$.
\end{enumerate}

To prove our lemma we will use a shortcut provided by the following characterization of matrices with positive definite \cite{bhatia2009positive} real part. Recall that if $A \in \C^{n\times n}$, its real and imaginary parts are $\Re A := \frac{1}{2}( A+A^\dagger)$ and $\Im A = \frac{1}{2\ii} (A-A^\dagger)$, respectively. 

\begin{theo}[see \cite{london1981note}] \label{theo:London}
Let $A \in \C^{n\times n}$. Then, $\Re A$ is positive definite if and only if 
\begin{equation}
A = T \begin{pmatrix} 1+\ii \alpha_1 & & \\ & \ddots &  \\ & & 1 + \ii \alpha_n \end{pmatrix} T^\dagger
\end{equation}
for some non-singular $T$ and $\alpha_1,\ldots,\alpha_n \in \R$.
\end{theo}

Note that, from condition 1 above,  
\begin{equation}
(\Id + 2 B_2^\dagger)(\Id + 2 B_2) = \Id \Rightarrow B_2 + B_2^\dagger + 2 (B_2^\dagger B_2) = 0,
\end{equation}
hence condition 1 is equivalent to $\Re B_2 = -B_2^\dagger B_2$. Recall that $A^\dagger A$ is positive definite for any $A \in \C^{n\times n}$, hence Theorem \ref{theo:London} can be applied to $-B_2$.

We recall the following parametrization of the $U(2)$ group, namely that
\begin{equation}
U(2) = \left\{ e^{\ii \theta} \begin{pmatrix}
\alpha & \beta \\ -\overline{\beta} & \overline{\alpha}
\end{pmatrix} : \theta \in [0,2\pi),~\alpha,\beta \in \C,~|\alpha|^2+|\beta|^2 = 1
\right\}. \label{eq:RepU2}
\end{equation}

\setcounter{lemma}{0}

\begin{lemma}
Let $B_1 \in \C^{2\times 2}$ be hermitian, with spectral decomposition $B_1 = V_1 D_1 V_1^\dagger$, and eigenvalues $d_1,d_2 \in [-1,1] \in \R$. Assume that $B_2 \in \C^{2\times 2}$ satisfies the conditions
\begin{enumerate}
\item $\Re B_2 = -B_2^\dagger B_2$,
\item $B_2^\dagger B_2 = \Id - B_1^2$.
\end{enumerate}
Let $\lambda_{i} = \sqrt{\smash[b]{1-d_{i}^2}}$, and $\eta_1 \in \{\eta_1^+, \eta_1^-\}$, $\eta_2 \in \{\eta_2^+, \eta_2^-\}$, with $\eta_i^+ \in [0,\pi/2]$, $\eta_i^- \in [-\pi/2,0]$, and such that $\sin\eta^\pm_{i} = \pm |d_{i}|$ for any $i \in \{1,2 \}$. Then, 
\begin{enumerate}
\item If $d_1^2 \neq d_2^2$ (non-degenerate case) then
\begin{equation}
B_2 = -V_1 \begin{pmatrix}
\lambda_{1} e^{\ii\eta_1} & 0 \\ 0 & \lambda_{2} e^{\ii\eta_2}
\end{pmatrix} V_1^\dagger.
\end{equation}
\item If $d_1^2 = d_2^2$ and $\eta_1 = \eta_2$, then $B_2 = -\lambda_1 e^{\ii\eta_1} \Id$.
\item If $d_1^2 = d_2^2$ and $\eta_1 = -\eta_2$, then
\begin{equation}
B_2 = -\lambda_1 K e^{\ii\eta_1\sigma_z} K^\dagger,
\end{equation}
for any $K \in U(2)$. We remark that any two $K_{1,2}$ such that $K_1 = K_2 K'$ for some $K'$ of the form $K' = \cos(\theta)\Id + \ii \sin(\theta)\sigma_z$, will give the same $B_2$.
\end{enumerate}

\begin{proof}
From Theorem \ref{theo:London}, and condition 1, $B_2$ can be written as $B_2 = -T_2 D_2 T_2^\dagger$ for some non-singular $T_2$ and $D_2 = \diag \{ 1+\ii\alpha_1,1+\ii\alpha_2 \}$, $\alpha_1,\alpha_2 \in \R$. Substitution into condition 1 gives 
\begin{equation}
T_2^\dagger T_2 = \begin{pmatrix}
(1+\alpha_1^2)^{-1} & 0 \\ 0 & (1+\alpha_2^2)^{-1}
\end{pmatrix}. \label{eq:ConstraintT2}
\end{equation}
Now, let the SVD of $T_2 = W \Sigma V^\dagger$. Then $T_2^\dagger T_2 = V \Sigma^2 V^\dagger$ and using \eqref{eq:ConstraintT2} it is easy to see, using the canonical decomposition of unitary matrices (cf. \eqref{eq:RepU2}), that we must have one of the following cases: 
\begin{enumerate}
\item If $\alpha_1^2 \neq \alpha_2^2$, then for some $\theta_1,\theta_2 \in [0,2\pi)$, either 
\begin{enumerate}
\item  $\Sigma = \begin{pmatrix} \frac{1}{\sqrt{1+\alpha_2^2}} & 0 \\ 0 & \frac{1}{\sqrt{1+\alpha_1^2}} \end{pmatrix}$ and $V = \begin{pmatrix} 0 & e^{\ii \theta_1} \\ e^{\ii \theta_2} & 0 \end{pmatrix}$. Hence, $B_2 = - W \begin{pmatrix}
\frac{1+\ii \alpha_2}{1+\alpha^2_2} & 0 \\ 0 & \frac{1+\ii \alpha_1}{1+\alpha^2_1}
\end{pmatrix} W^\dagger $. 

\item $\Sigma = \begin{pmatrix} \frac{1}{\sqrt{1+\alpha_1^2}} & 0 \\ 0 & \frac{1}{\sqrt{1+\alpha_2^2}} \end{pmatrix}$ and $V = \begin{pmatrix} e^{\ii \theta_1} & 0 \\ 0 & e^{\ii \theta_2} \end{pmatrix}$. Hence, $B_2 = - W \begin{pmatrix}
\frac{1+\ii \alpha_1}{1+\alpha^2_1} & 0 \\ 0 & \frac{1+\ii \alpha_2}{1+\alpha^2_2}
\end{pmatrix} W^\dagger $.  

Note that in either case, we can write $B_2 = - W \diag \left\{ \frac{1+\ii \alpha_{\sigma(1)}}{1+\alpha^2_{\sigma(1)}}, \frac{1+\ii \alpha_{\sigma(2)}}{1+\alpha^2_{\sigma(2)}} \right\}  W^\dagger $, where $\sigma : \{ 1,2\} \to \{ 1,2\} $ is a permutation. It is easy to check that condition 1 is indeed satisfied. Next, substitution into condition 2 gives
\begin{equation}
\begin{pmatrix}
1-d_1^2 & 0 \\ 0 & 1-d_2^2
\end{pmatrix} = K \begin{pmatrix} \frac{1}{1+\alpha_{\sigma(1)}^2} & 0 \\ 0 & \frac{1}{1+\alpha_{\sigma(2)}^2} \end{pmatrix} 
K^\dagger,
\end{equation}
with $K := V_1^\dagger W \in U(2)$. Again, we use the canonical form, cf. \eqref{eq:RepU2}, to find that $K$ is diagonal or antidiagonal, with two independent phases. Introducing back $W$ into $B_2$, in either case we have $\alpha_{i}^2 = \frac{d_{i}^2}{1-d_{i}^2}$, $i=1,2$, hence $\alpha_{i} = \pm \frac{|d_i|}{\sqrt{1-d_i^2}}$, so
\begin{equation}
\frac{1+\ii \alpha_{i}}{1+\alpha^2_{i}} = \sqrt{\smash[b]{1-d_{i}^2}}\left( \sqrt{\smash[b]{1-d_{i}^2}} \pm \ii |d_{i}| \right) = \sqrt{\smash[b]{1-d_{i}^2}} e^{\ii \eta^\pm_{i}}, \label{eq:AppBSimplAlpha}
\end{equation}
provided that $\cos \eta^\pm_{i} = \sqrt{\smash[b]{1-d_{i}^2}}$, and $\sin \eta^\pm_{i} = \pm |d_{i}|$. This proves part 1.

\end{enumerate}
\item If $\alpha_1^2 =  \alpha_2^2$, then $\Sigma = \frac{1}{\sqrt{1+\alpha_1^2}} \Id_2$, and $V \in U(2)$ is arbitrary. Thus, we can write $
B_2 = -\frac{1}{1+\alpha_1^2} K \begin{pmatrix}
1+\ii \alpha_1 & 0 \\ 0 & 1\pm \ii \alpha_1
\end{pmatrix} K^\dagger $ for some $K \in U(2)$. Substitution into condition 2, gives $\alpha_1^2 = d_1^2 / (1-d_1^2)$, and proceeding as in \eqref{eq:AppBSimplAlpha}, we obtain the claim.

\end{enumerate}

\end{proof}

\end{lemma}

Next we characterize $B_4$. We assume that the relevant constraints fron \eqref{eq:condBDiagonal}-\eqref{eq:condBAntiDiagonal2} are satisfied, namely $B_1^2 + B_2^\dagger B_2 = \Id_2$ and that $B_4^2 + B_2 B_2^\dagger = \Id_2$. 

\begin{lemma}
In the hypothesis of above, 
\begin{enumerate}
\item If $d_1^2 \neq d_2^2$ (non-degenerate case), then $B_4=-B_1$.

\item If $d_1 = d_2$, then $B_4 = -B_1$.
\item If $d_1 = -d_2$, then
\begin{equation}
B_4= d_1 K \sigma_z K^\dagger, 
\end{equation}
for any $K \in U(2)$. We remark that any two $K_{1,2} \in U(2)$ such that $K_1 = K_2 K'$ for some $K'$ of the form $K' = \cos(\theta)\Id + \ii \sin(\theta)\sigma_z$, will give the same $B_4$.
\end{enumerate}

\begin{proof}
From \eqref{eq:condUB}, $B_2$ is normal, then from \eqref{eq:condBDiagonal} and \eqref{eq:condBDiagonal2} we have that $B_1^2=B_4^2$. Since $B_4$ is hermitian consider its spectral decomposition, $B_4=W D_4 W^\dagger$, $W \in U(2)$. Then, $D_4^2 = K D_1^2 K^\dagger$, where $K := W^\dagger V_1$. Using the canonical form \eqref{eq:RepU2}, we find that
\begin{enumerate}
\item If $d_1^2 \neq d_2^2$, then $K$ is either diagonal or anti-diagonal, with arbitrary phases. In either case we obtain $B_4 = V_1\diag\{ \pm d_1,\pm d_2 \} V_1^\dagger$, but since we must have $\trace B_4=-\trace B_1$, we shall take $-d_1$, $-d_2$. Hence, $B_4=-B_1$.
\item If $d_1^2 = d_2^2$, then $K \in U(2)$ is arbitrary, and we have $B_4 = V_1 K^\dagger \diag\{ \pm d_1,\pm d_1 \} K V_1^\dagger$. If $d_1=d_2$, then $\trace B_4 = -2d_1$, so $B_4 = -d_1 \Id_2 = -B_1$. If $d_1=-d_2$, then $\trace B_4=0$, and we can take $d_1,-d_1$ or $-d_1, d_1$. 
\end{enumerate}
\end{proof}
\end{lemma}

\end{widetext}

\end{document}